# Ranking Critical Tools in the Implementation of Lean Six Sigma as an Integrated Management System in Portugal

David Ferreira and Pedro F. Cunha

Instituto Politécnico de Setúbal - Escola Superior de Tecnologia de Setúbal,
Setúbal, Portugal
david.neto@estudantes.ips.pt, pedro.cunha@estsetubal.ips.pt

**Abstract.**. Lean Six Sigma (LSS) is a comprehensive and powerful strategy for processes improvement and products. There is a cornucopia of tools for its implementation and 37 among them were selected to carry out an evaluation based on three factors, namely: Frequency of use of the tool; Difficulty in implementing; Importance and impact of the tool in the implementation of LSS. An online survey was conducted with Portuguese consultants and it included questions on the profile, and the companies they worked, as well as the degree of impact of the tools used. Consultants were asked to choose ten tools, ranking them in order of importance. The frequencies with which each tool had been cited were counted. A procedure was then developed to identify the know-how of consultants to establish a ranking of LSS tools. It was created an ordering list of tools, which emphasized in: Honshin Kanri, VOC, VSM. The results presented are particularly relevant when is considered the importance of understanding the requirements for a successful implementation of Lean Six Sigma management system in the organizations.

## 1 Introduction

Currently, manufacturing industries increasingly seek to achieve excellence in their operations and production activities to become competitive in the actual global business environment. With this aim, companies are challenged to implement systems or models, which comprises the right set of methods and tools, that helps them to achieve operation excellence. Lean Manufacturing, as a management philosophy or management model, has it origin in the years 50s, with the purpose to increase competitiveness, when Taiichi Ohno and Shigeo Shingo developed a new production system for Toyota, whose objective was focused on identifying and eliminating waste and increasing the speed of product delivery [8]. Six Sigma gain its expression, as we know today, in the years 80s when Motorola Inc. started to use tools that reduce variation and defects to deliver products and services to customers that meet their expectations [20]. It is a comprehensive and flexible system for achieving, supporting and maximizing business success: through understanding customer needs and statistical analysis to manage, and reinvent business processes [10]. The integration of Lean Manufacturing and Six Sigma is called Lean Six Sigma (LSS) and constitutes a comprehensive, powerful and effective strategy for process improvement and creation of high-quality products [11]. Much has been said about the comparison between Lean and Six Sigma [6], [9], however, another aspect of great impact is the synergism of LSS with Industry 4.0 [15], [18].

This paper intends to evaluate Lean Six Sigma management tools, in terms of frequency of use, degree of implementation difficulty and impact through a Decision

Matrix. In the end, by extracting the opinions of experts, a ranking of use of the tools to support the LSS is presented.

## 2    Online Survey

Research was conducted to verify the number of agents or consultants that are dealing with LSS management system in Portugal, and a number of 21 consulting firms and 50 consultants were compiled. Between them, there are a variety of academic degree and experience [19].

In order to obtain a quantitative view on the implementation of LSS in Portugal, a questionnaire was structured, created on the Qualtrics platform [13]. This questionnaire consisted of 22 questions and was divided into the following topics: Consultant profile; Profile of the industries operated, Tools used in Lean Six Sigma (Frequency of use, Level of difficulty, Degree of impact). The survey was sent to the 50 professionals found via LinkedIn and/or e-mail, and a total of 19 responses were obtained. However, four of them were rather incomplete and were discarded. Regarding the profile of the 15 consultants who responded to the survey, it was identified that almost 80% of their academic training is in production or industrial engineering, with chemical engineering in third place with 12.5%. Regarding the time of experience in the market working as LSS analysts, the results obtained from the survey showed almost 70% with a period of less than 10 years of experience, while the most experienced totaled 31.2%. To act as an LSS consultant [16], a Black Belt (or higher) certification was expected. It was noted that more than half (54%) have such a certification. Regarding the geographic location and coverage of consultant's activities in Portuguese manufacturing industries, it was observed in the sample that the majority worked in the district of Porto (75%), followed by Aveiro (67%) and Lisbon (58%). This result is in agreement with the information found at [5], [7], which indicates a greater concentration of industries in the north of the country [12]. Regarding the main sectors of the manufacturing industry that consultants provide or have provided services in Portugal, the automotive industry can be highlighted with 60% [14]. More than half (53%) also worked in the food, machinery, and metalworking sectors.

## 3    Ranking of LSS-tools in Portugal

In the survey, consultants were asked to evaluate 37 Lean Six Sigma tools, considering three factors for analysis:

- Frequency of tool use;
- Difficulty and/or resistance in implementing the tool;
- Importance and impact of the tool in the implementation of LSS.

The following subsections address these different aspects of the LSS tools, according to the perspective or experience of the analysts who joined the survey.

### 3.1    Usability Ranking of LSS tools

For the frequency of the tools, a Likert 1-5 scale was used [1]. The arithmetic mean was calculated for the degree of frequency of each tool and ordered from the most used to the least used (see Fig. 1).

Among the tools, the one that can be considered "always" applied (average score > 4) is the PDCA Cycle, Gemba Walk, Kaizen, 5S, in blue, covering about 15% of the investigated tools. It followed by others "almost always" applied, such as the Ishikawa, A3 Report, among others (in green color, totaling 30% of all tools). In contrast, the class of very least used tools (in orange) are, in order of least use: TOC,

Harada Method, BSC, QRQC, Benchmarking and FTA (representing about 15% of the pool of tools). Here, just one aspect is concerned: the frequency of use reported in the literature, without considering the difficulty in implementing the tool nor its impact on the implementation of the LSS. The next ranking was built by the same token, but now taking only in account the consultants' perception. Kaizen, PDCA and VSM are the highlight in use this round. Gemba Walk, 5 Whys, 5S, Hoshin Karin and Spaghetti Diagram follow. The top 10 in each case roughly overlap. A mention only for VSM (Value stream mapping) who played a prominent role just in the latter case.

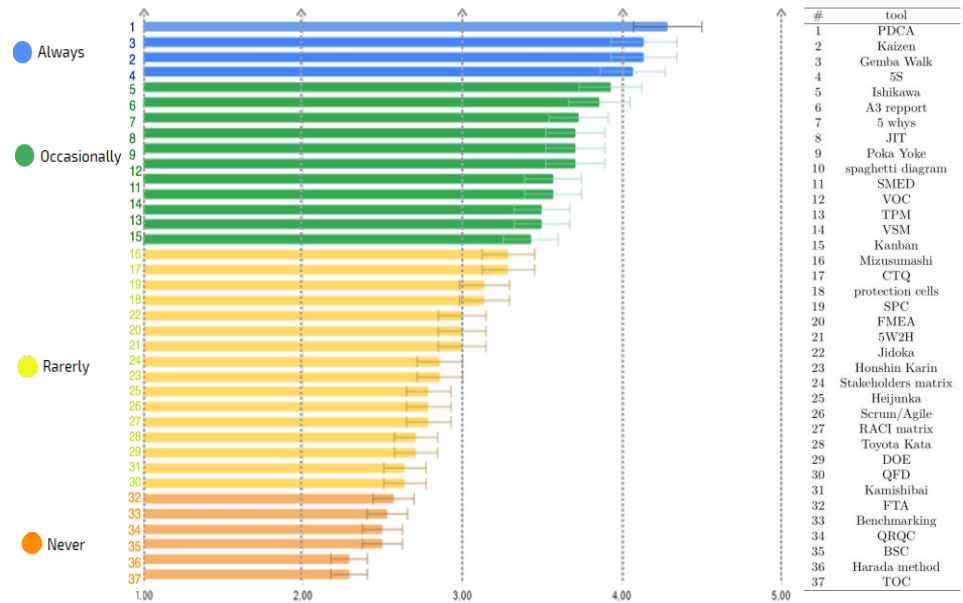

**Fig. 1.** Frequency of use of tools according to survey.

### 3.2  Resistance Ranking in the Application of LSS-tools

In the case of difficulty and/or resistance in the application of each tool, a Likert scale that ranged from 1 to 3 was applied (see Fig. 2).

High was: FMEA, TPM and 5S. Now, the FMEA [2] is applied in about 12 stages that include diagnosis, evaluation, definition of strategies and monitoring for the prevention/correction of failures by the teams; it serves to create a culture for the search for "0 defects", or rather, with 99.99966% quality. TPM is supported by 8 pillars [4], with a general approach to preventive maintenance in order to achieve sustainability and "perfect production". Finally, 5S (one of Lean Six Sigma's finest tools), which is an approach that encompasses 5 steps, requires systematic planning that involves several factors such as: optimizing productivity, reducing expenses, promoting well-being and safety. About the tools with a lower degree of resistance or implementation difficulty, the following can be highlighted (in increasing order of difficulty): Ishikawa Diagram, Spaghetti Diagram, CTQ and 5 Whys. What these tools have in common is that they are faster, simpler and straightforward to apply for short-term spot investigations.

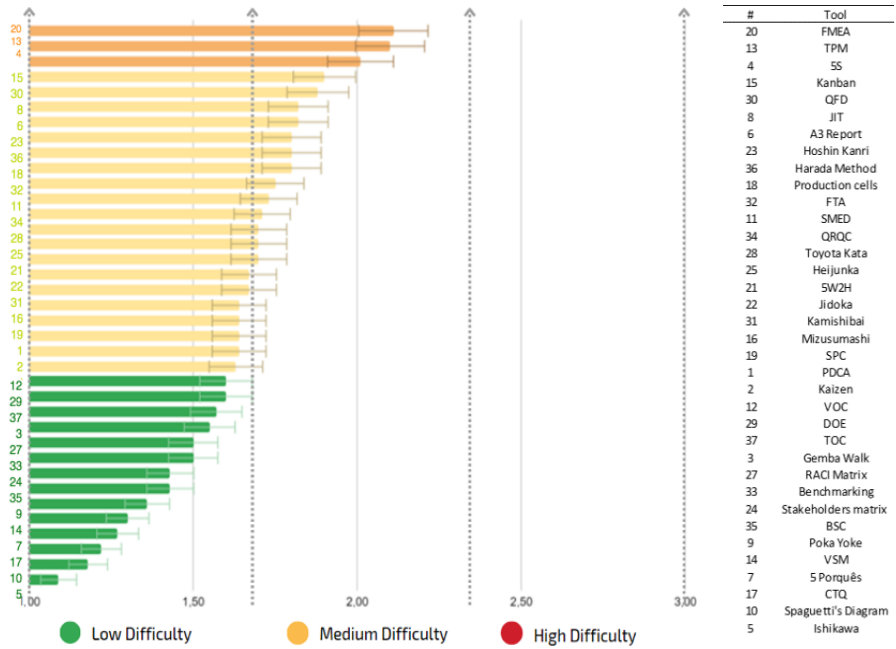

**Fig. 2.** Perception of difficulty in applying the tool according to Portuguese consultants.

The last question of the survey asked consultants to choose 10 tools with the greatest impact on LSS implementation according to their experience and professional practices, ranking them in ascending order of importance.

It is necessary to consider the weight of the position of each tool in the responses obtained, and for that, a central measure of the opinions of different analysts about each tool is needed. In Statistics, there are several ways to measure central tendency. The mean is often adopted for symmetrical distributions with a low number of outliers. The median is generally preferred to return to central tendency in the case of skewed distributions with high variability. In the case of consultants' perception, consistency and unanimity is not to be expected, so the average does not seem to be the best choice in the evaluation of the "average opinion". It is relatively frequent some conflicting opinions, due to different views and experiences, increasing the dispersion and asymmetry (see Table 1).

**Table 1.** Ranking of Tool's Impact according to survey.

| Tool | F | F% | Position | Mean | Median | Final |
|---|---|---|---|---|---|---|
| Honshin Kanri | 5 | 0,45 | 1º, 1º, 1º, 9º, 4º | 3,2 | 1º | 1º |
| VOC | 4 | 0,36 | 1º, 1º, 8º, 4º | 3,5 | 3º | 2º |
| VSM | 8 | 0,73 | 5º, 2º, 4º, 4º, 5º, 3º, 3º, 3º | 3,6 | 3,5º | 3º |
| 5S | 5 | 0,45 | 2º, 6º, 7º, 4º, 2º | 4,2 | 4º | |
| Esparguete | 5 | 0,45 | 4º, 4º, 5º, 6º, 2º | 4,2 | 4º | 4º |
| Kanban | 4 | 0,36 | 1º, 7º, 7º, 1º | 4 | 4º | |
| PDCA | 8 | 0,73 | 6º, 3º, 3º, 8º, 3º, 2º, 8º, 9º | 5,3 | 4,5º | 5º |
| 5 Porquês | 6 | 0,54 | 9º, 3º, 5º, 6º, 3º, 6º | 5,3 | 5,5º | 6º |
| Gemba Walk | 6 | 0,54 | 10º, 5º, 6º, 2º, 6º, 7º | 6 | 6º | 7º |
| Ishikawa | 4 | 0,36 | 7º, 4º, 7º, 10º | 7 | 7º | 8º |
| Kaizen | 9 | 0,82 | 4º, 5º, 2º, 8º, 9º, 8º, 8º, 1º, 8º | 5,9 | 8º | 10º |

By selecting the most voted and calculating their respective medians, it was possible to rank recommended tools, with emphasis on: (1st) Honshin Kanri, (2nd) VOC, (3rd) VSM, (4th) 5S, Spaghetti Diagram and Kanban (5th). Although Hoshin-Kanri was not identified as a high-frequency tool in previous assessments, it appeared in the ranking in first place, given its strong impact on the definition and deployment of strategic goals at all levels of the organization that guide improvement projects

## 4  Decision Matrix for LSS Tools

To facilitate the analysis of how these tools were evaluated, an adapted Decision Matrix was developed. A Cartesian plane was considered, where the tool application frequency was represented in the *Y*-coordinate and the Tool application difficulty in the *X*-coordinate. The 2D-points were plotted, and a characterization was proposed as follows:

1. "Quick Wins" tools: they have a high frequency of use and low implementation difficulty, they can be applied quickly with low resistance;
2. "Big Projects" tools: they have high frequency and high difficulty, they are tools that need a longer period to apply;
3. "Plan B" tools: they have a low frequency of use and low difficulty and can be considered secondary tools, since even with little resistance or degree of complexity, they have a lower rate of use, applicability or impact;
4. "Low use" tools: have low frequency of use and high difficulty in implementation; tools that are not normally used.

In the Decision Matrix, when considering the tools as the most suitable for application would be in the "Quick Wins" quadrant; while the less indicated tools in "Low Use". So, as to map the tools in the decision matrix, the values obtained from the averages of frequencies and the degree of difficulty were tabulated and used as coordinates to plot on Fig. 3 (color code: blue=1st quadrant, green=2nd quadrant, orange=3rd quadrant).

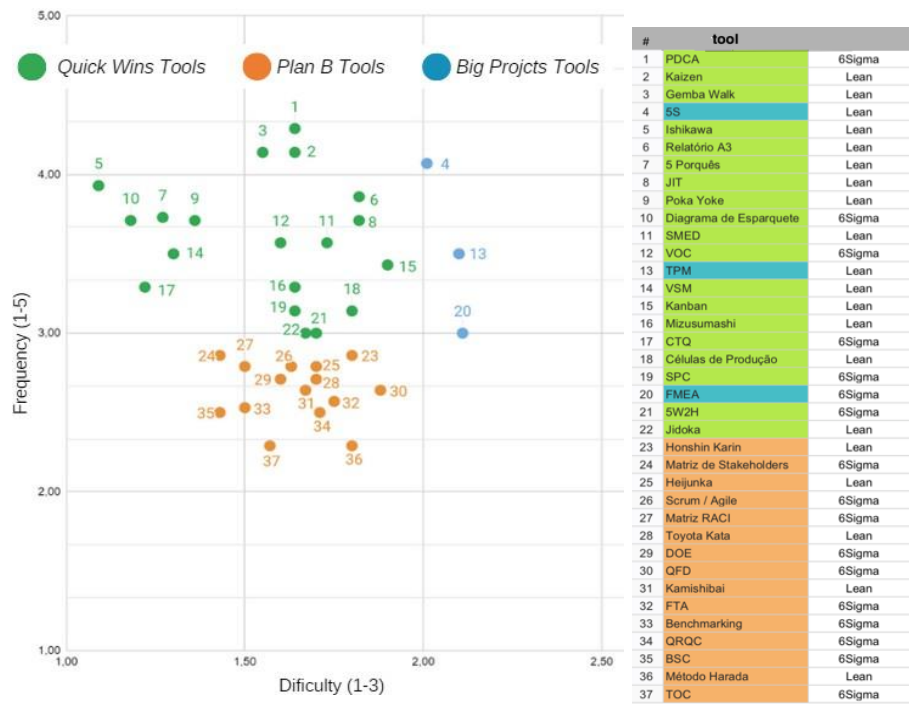

**Fig. 3.** Scatterplot of LSS-tools in the decision matrix in consultants' perspective.

What is observed is, according to the consultants' perspective, that the highest concentration of tools (92% in total) is classified in the first and third quadrants, Quick Wins (51% of the total) and Plan B (41% of the total), respectively. Regarding those considered as Quick Wins, the use of PDCA, Kaizen, Gemba Walk, Ishikawa Diagram, A3 Report stands out. Once again is observed, now more clearly that most of the Lean tools were considered quick wins, while most Six Sigma tools were classified as "Plan B". Only three of the tools were classified as "Big Projects", namely: 5S, TPM and FMEA. It is also noteworthy that this result reflects the perception and experience of the analysts in the sample and that, its choice and use depend on the needs of the company and the project. When verifying the result, Hoshin-Kanri stands out in first place, with an average degree of use. It is a strategic tool for the deployment of goals related to the vision and organizational objectives at all levels and departments (horizontal and vertical). In this way, consultants are concerned about the importance of a strategic vision to guide and guarantee the successful implementation of LSS. In second place appears the VOC, despite low reliability, demonstrating the importance of the "Voice of the Customer" (whether internal or external) in the survey of needs and parameters for improvement opportunities, as well as in the translation of CTQs (critical factors) and in the establishment of key indicators to guide projects. The choice of VSM and 5S tools have good reliability and were ranked, respectively, in $3^{rd}$ and $4^{th}$ place. While the VSM maps the value stream chain, which allows identifying waste and maximizing value for the customer, the 5S (*Seiri*, *Seiton*, *Seiso*, *Seiketsu*, *Shitsuke*) is a more complete system for handling workplace organization. On the other hand, despite the Kaizen tool being the most voted (highest occurrence), its placement appears in the last position in the top-10 ranking, which indicates a great agreement between the consultants.

## 5   Concluding Remarks

In this paper, we sought to rank the tools used in the implementation of the LSS, using the perception of specialized consultants. In order to evaluate these tools, a Decision Matrix was built that took into account the frequency of application of the tool and its degree of difficulty / resistance in the application. Most Lean tools were considered quick wins, while most Six Sigma tools were classified as Plan B. A procedure was developed to establish a ranking of tools according to the Portuguese consultants who responded to the survey, considering their perceptions and practical experiences. The objective was to raise the critical tools in the LSS implementation.

By selecting the most voted and calculating their respective medians, it was possible to rank recommended tools, with emphasis on: ($1^{st}$) Honshin Kanri, ($2^{nd}$) VOC, ($3^{rd}$) VSM, ($4^{th}$) 5S, Spaghetti Diagram and Kanban ($5^{th}$). Although Hoshin-Kanri was not identified as a high-frequency tool in previous assessments, it appeared in the ranking in first place, given its strong impact on the definition and deployment of strategic goals at all levels of the organization that guide improvement projects. In second place comes the VOC, which emphasizes the importance of the "Voice of the Customer" in the survey of needs for opportunities for continuously im- proving productivity and product quality. A value-stream map (VSM) is often created to reflect the actual current operation status in the production and appears as valuable tool and had a high level of recommendation.

With this characterization of tools, being used in the implementation of Lean Six Sigma, it was given an up-to-date overview of the main issues related to the implementation of continuous improvement in the Portuguese industry. This study it will be particularly valuable when combined with the knowledge about the barriers for Lean Six Sigma implementation. Mainly, it will help to establish the requirements or to define a roadmap for a successful integrated management system accomplishment.